\documentclass[aps,prl,preprint,superscriptaddress]{revtex4-1}
\usepackage[usenames]{color}
\usepackage{epsf}
\usepackage{epsfig}
\usepackage{graphicx}
\usepackage{latexsym}
\usepackage{bm}
\begin{document}
~\\
published in: Phys. Rev. Lett. 104, 133401 (2010)
~\\
\title{Interatomic Coulombic Decay following Photoionization of the Helium Dimer:\\
Observation of Vibrational Structure }

\author{
T. Havermeier$^{1}$, T. Jahnke$^{1}$, K. Kreidi$^{1}$, R.
Wallauer$^{1}$, S. Voss$^{1}$, M. Sch\"offler$^{1}$,
S.Sch\"ossler$^{1}$,\\ L. Foucar$^{1}$, N. Neumann$^{1}$, J.
Titze$^{1}$, H. Sann$^{1}$, M. K\"uhnel$^{1}$, J.
Voigtsberger$^{1}$, J. H. Morilla$^{3}$, W. Sch\"ollkopf$^{2}$, H.
Schmidt-B\"ocking$^{1}$, R. E. Grisenti$^{1,4}$ and R.
D\"orner$^{}$}


\address{
Institut f\"ur Kernphysik, J.~W.~Goethe Universit\"at, Max-von-Laue-Str.1, 60438 Frankfurt, Germany \\
$^2$Fritz-Haber-Institut der Max-Planck-Gesellschaft, Faradayweg 4-6, 14195 Berlin, Germany\\
$^3$Instituto de Estructura de la Materia, CSIC Serrano 121, 28006
Madrid, Spain\\
$^4$GSI Helmholtzzentrum f\"ur Schwerionenforschung GmbH Planckstr.
1, 64291 Darmstadt, Germany}

\begin{abstract}
Using synchrotron radiation we simultaneously ionize and excite one
helium atom of a helium dimer (He$_2$) in a shakeup process. The
populated states of the dimer ion (i.e. He$^{*+}(n=2,3…)-$He) are
found to deexcite via interatomic coulombic decay. This leads to the
emission of a second electron from the neutral site and a subsequent
coulomb explosion. In this letter we present a measurement of the
momenta of fragments that are created during this reaction. The
electron energy distribution and the kinetic energy release of the
two He$^+$ ions show pronounced oscillations which we attribute to
the structure of the vibrational wave function of the dimer ion.
\end{abstract}
\maketitle

Excited hydrogen-like atoms and ions can deexcite only by emission
of a photon. If however these excited particles are put into the
vicinity of other atoms, the excitation energy can in principle be
transferred to the neighbor, where it may lead to emission of an
electron. This two-center energy transfer process is known as
interatomic coulombic decay (ICD). It was first predicted by
Cederbaum and coworkers for molecular clusters
\cite{Cederbaum1997PRL}. Today it is well established also
experimentally for inner valence excitation of many electron systems
such as van-der-Waals clusters containing Ne, Ar and Xe (see e.g.
\cite{Marburger2003PRL,Jahnke2003PRL,Morishita2006PRL,Lablanquie2007JCP})
and water clusters \cite{Jahnke2009water}.

In the present experiment we demonstrate the existence of ICD in the
most fundamental system in which it can occur: excited
He$^{*+}(n=2...)$ with a van-der-Waals-bound neighboring neutral
helium atom. We observe, that different from all the previously
considered systems, the energy distribution of the low energy
electron emitted via ICD from He-He exhibits an oscillatory
structure. Furthermore the kinetic energy of the ionic fragments
(KER) reveals that ICD occurs at interatomic distances up to
$\approx$ 12 a.u. implying that no overlap of the electronic wave
function is necessary for ICD to occur.

With a binding energy of only 95 neV, He$_2$ is the most weakly
bound naturally occurring system
\cite{Schoellkopf1994Science,Grisenti2000PRL}. The delicate
interplay of zero point motion and weakness of the He-He
van-der-Waals potential results in an extremely delocalized nuclear
wave function (see Figure 1), which is qualitatively different from
all other known clusters: the mean value of the internuclear
distance is $\approx 52\,{\rm \AA}$ \cite{Grisenti2000PRL,Gentry},
which is off the scale of Figure 1. The wave function extends from
$\approx$ 5 a.u. to several 100 a.u. Such a delocalized ensemble is
an ideal starting point for studies of ICD. The coulomb explosion
following ICD allows to watch the decay of this ensemble. Measuring
the kinetic energy release (KER) and the direction of the fragments
allows to detect the internuclear distance at which ICD occurred for
each individual event.

\begin{figure}[ht]
  \begin{center}
  \includegraphics[width=9 cm]{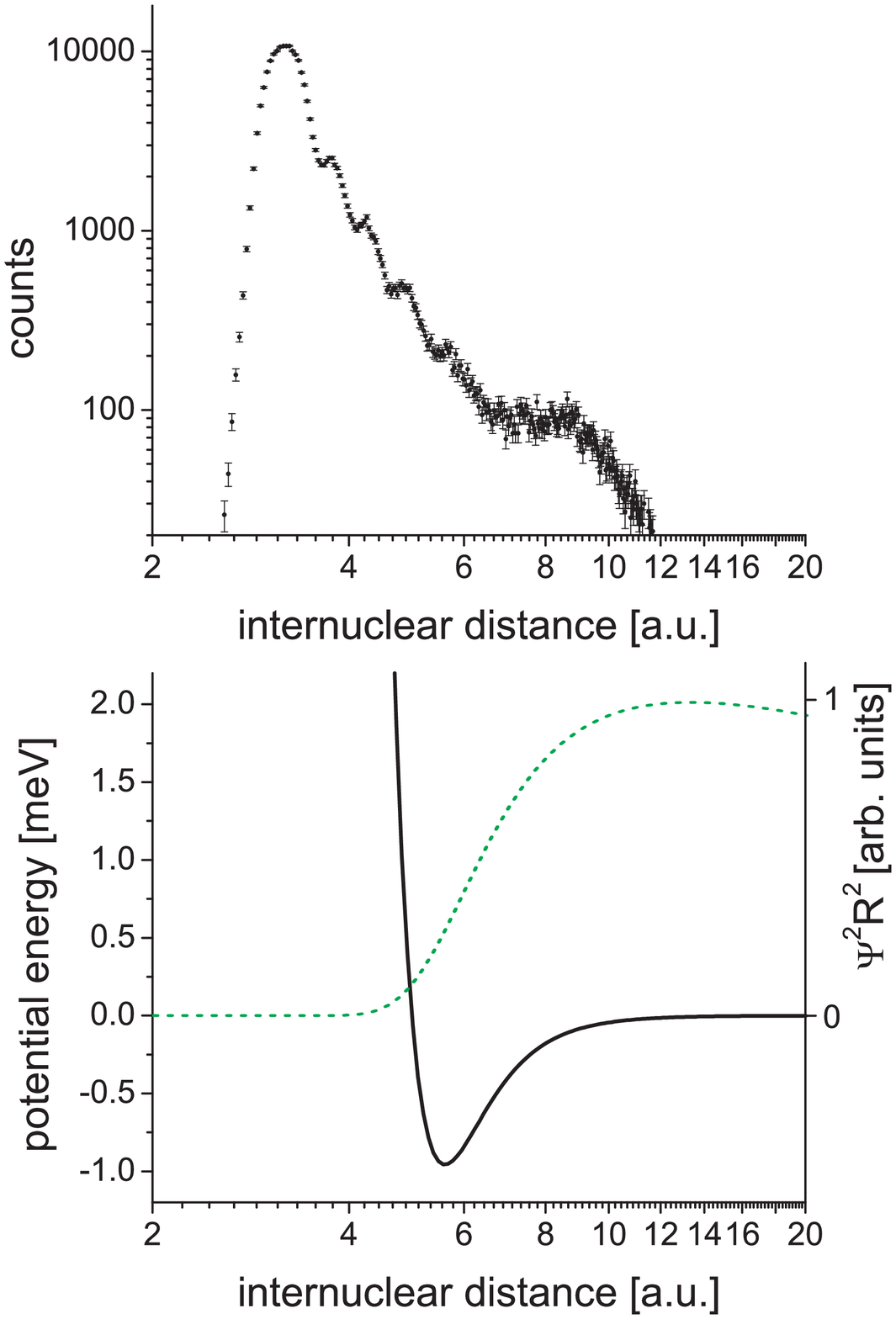}
  \caption{Bottom: He dimer potential (left scale)(from \cite{Tang1995PRL}).
  dotted green line in right axis: $\psi^2 R^2$ (from \cite{Luo1993JCPb}).
  The expectation value of internuclear distance is 52 ${\AA}$ \cite{Grisenti2000PRL},
  which is off scale. Upper panel: measured internuclear
  distance at instant of ICD from $He^{*+}(n=2)-He \rightarrow He^+ + He^+ + e_{ICD}$.
  Calculated from KER (see fig. 3) at a photon energy of
  68.86 eV using the reflection approximation R=1/KER}
  \end{center}
\end{figure}

The experiment was performed at beamline UE112PGM2 at BESSY using
the COLTRIMS technique
\cite{doerner00pr,Ullrich03RPP,Jahnke04JESRP}. We create helium
dimers by expanding He gas through a 5 $\mu$m nozzle cooled to 18 K
by a continuous flow cryostat. A driving pressure of 1.8 bar and a
pressure of 1.2$\times$10$^{-4}$ mbar at the low pressure side of
the nozzle yielded a dimer fraction of 1-2 \% in the gas beam. This
fraction has been determined using diffraction at microstructure
gratings as performed in \cite{Schoellkopf1994Science}.  For the
given conditions the fraction of trimers and larger clusters was
below 0.2 \% of the monomers. 10 mm above the nozzle the beam
entered a scattering chamber through a 0.3 mm skimmer. The gas beam
was intersected with a linearly polarized photon beam in the center
of a homogenous electric field region of a COLTRIMS spectrometer.
The electric field and a parallel homogenous magnetic field of 10 G
guided the electrons and ions towards two microchannel plate
detectors (80 mm active diameter) with delay-line position readout
\cite{Jagutzki2002NIMP}. From the measured positions of impact and
times-of-flight of ions and electrons their respective momentum
vectors and charge to mass ratios are obtained. The back-to-back
emission of the two He$^+$ ions provides a unique signature allowing
for an almost perfect suppression of the huge background of ions and
electrons from ionization of the monomers. It also allows to detect
possible contaminations resulting from the fragmentation of larger
clusters. For all data shown in this paper we have selected only
those events where two He$^+$ ions are emitted back-to-back. At an
ion rate of about 10 kHz we observed a rate of about 7 Hz for these
He$^+$ ion pairs. The ion momenta were calibrated using coulomb
explosion of N$_2$ at 77.86 eV photon energy which leads to a narrow
peak at a KER of 10.32 eV \cite{Lundquist1996,Weber2001}.

\begin{figure}[ht]
  \begin{center}
  \includegraphics[width=8 cm]{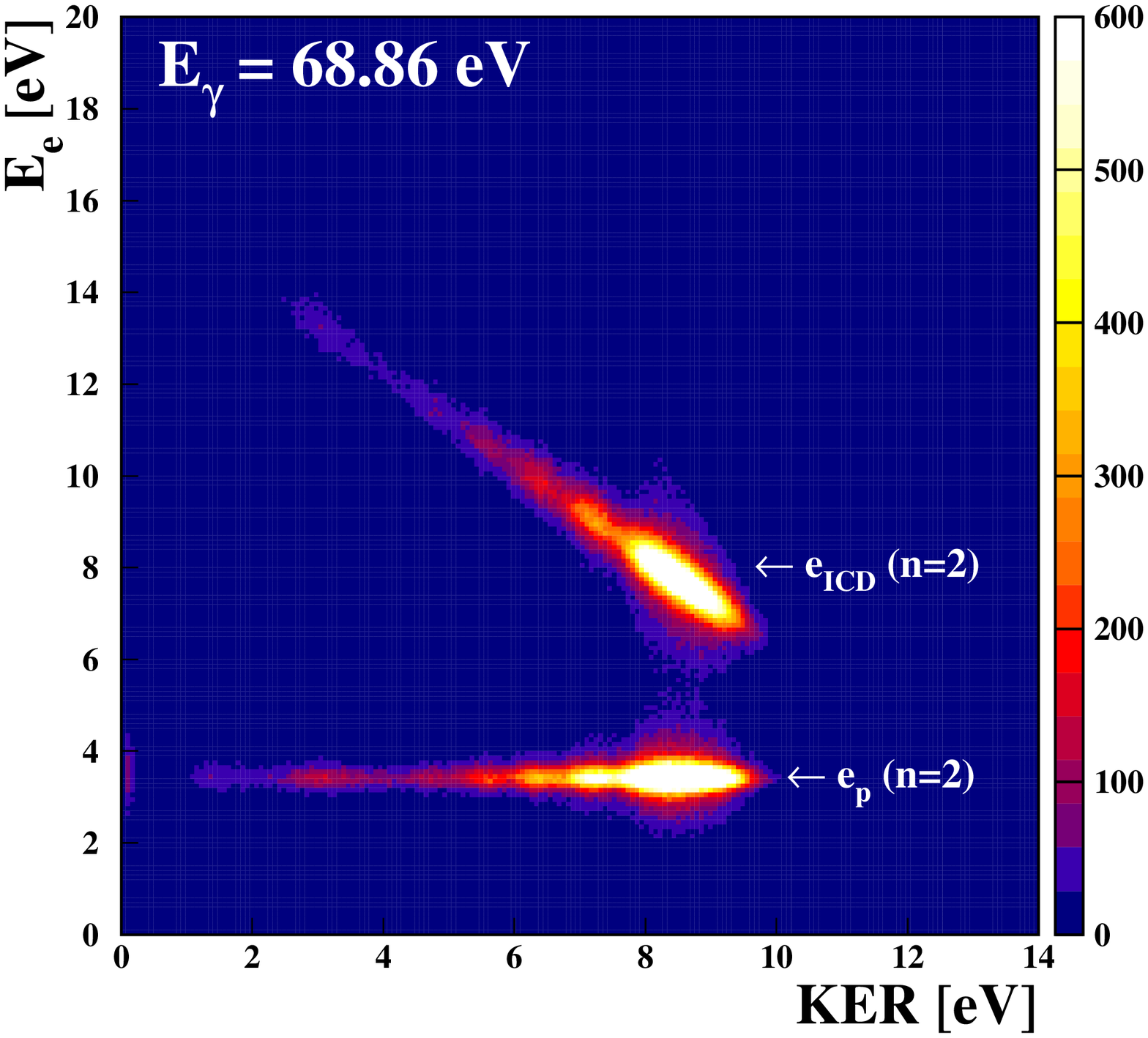}
  \includegraphics[width=8 cm]{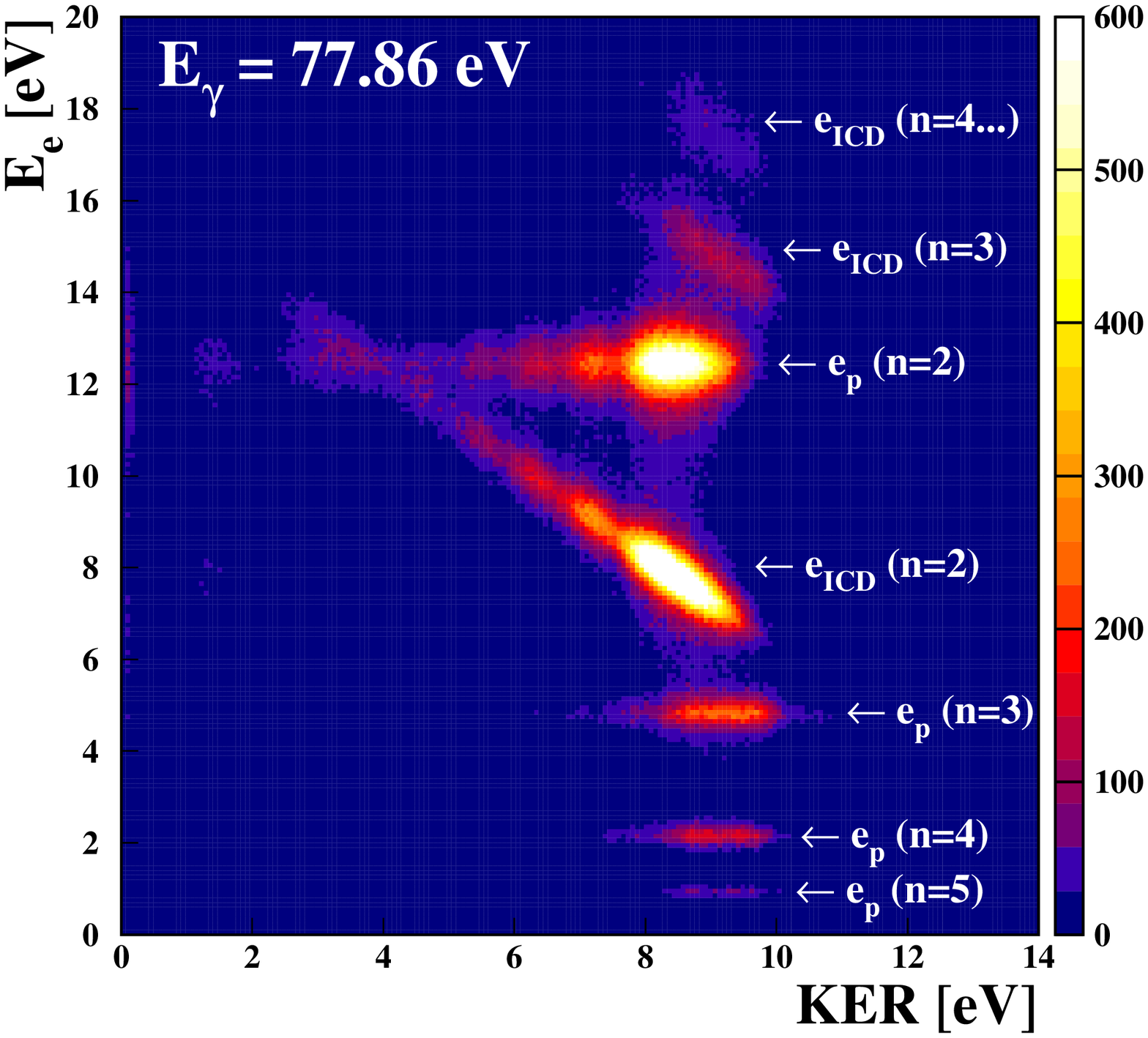}
  \caption{Kinetic energy release of the He$^{+}$ - He$^{+}$ fragments versus the energy
  of one of the two electrons from photoionization excitation of He$_2$ at photon energies of
  68.86 and 77.86 eV. The arrow indicate the expected position of photoelectron (horizontal line)
  and the corresponding ICD electron (diagonal lines) for excitation to the intermediate state He$^+$(n)-He.}
  \end{center}
\end{figure}

Figure 2 shows the energy of one of the detected electrons versus
the KER at photon energies of E$_\gamma$=68.68 eV and 77.86 eV. The
creation of He$^{*+}$(n) in the excited state of principal
quantumnumber n by photoionization plus simultaneous excitation
results in photoelectrons of an energy E$_{ephoto}$(n)= E$_\gamma$ -
24.59eV -(54.42eV- 13.6eV $\frac{4}{n^2})$ (as depicted by  arrows
in Fig. 2). In addition to these horizontal lines of the satellite
photoelectrons Figure 2 shows also events along diagonal lines. A
diagonal indicates a constant sum of the two quantities plotted on
the x- and y-axis i.e. a constant sum energy of the electrons and
ions observed. These lines are characteristic for ICD
\cite{Jahnke2003PRL}: They appear for the case of a decay of an
intermediate state with discrete energy that partitions its energy
among the kinetic energy of the electron emitted during the decay
and the KER of the ions. The positions in energy expected for the
decay of He$^{*+}$(n)-He $\rightarrow$ He$^+$(n=1)-He$^+$(n=1) +
e$^-$ are indicated for the n=2,3,4 in Figure 2. For those events
where we detect both electrons in coincidence we have confirmed that
each photoelectron is accompanied by its matching ICD electron.

\begin{figure}[ht]
  \begin{center}
  \includegraphics[width=10 cm]{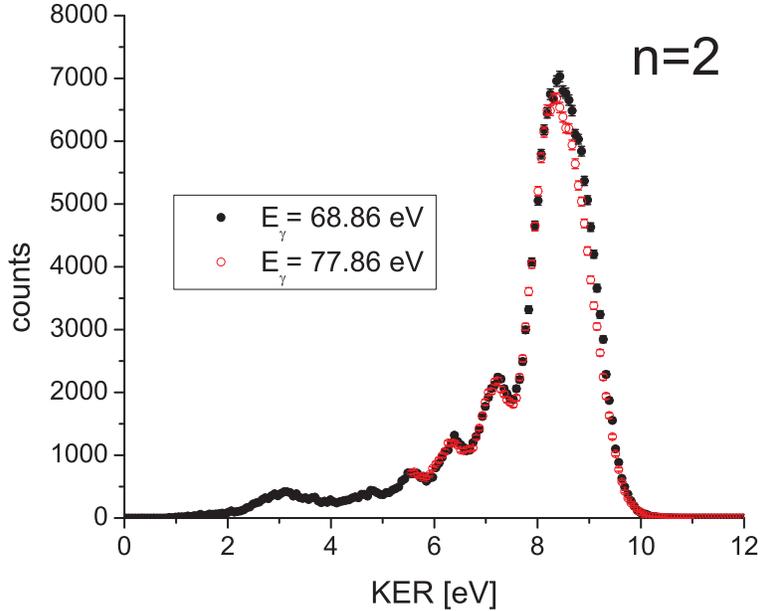}
  \caption{Kinetic energy release of the He$^{+}$ - He$^{+}$ fragments for intermediates
  excited states with principal quantum n=2 (He$^{*+}$(n=2) - He).
  Data are obtained by projecting the respective diagonal
  line in Figure 2 onto the KER axis.}
  \end{center}
\end{figure}

\begin{figure}[ht]
  \begin{center}
  \includegraphics[width=12 cm]{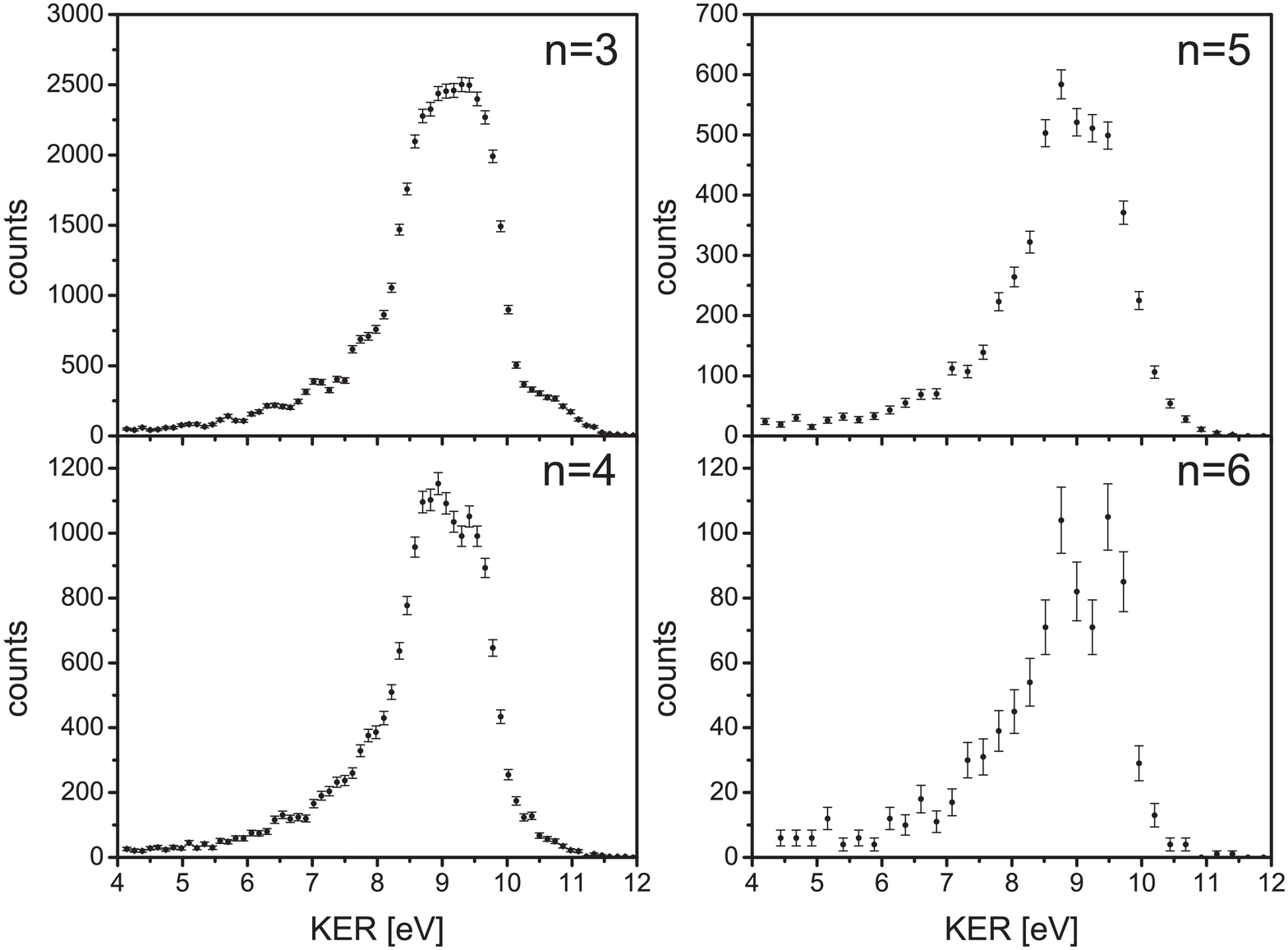}
  \caption{Distribution of the kinetic energy release for intermediates
  excited states (n) at a photon energy of 77.86 eV. The principal quantum
  number n as given in the figure.}
  \end{center}
\end{figure}

The measured KER for the different n is shown in Figure 3 and 4.
These spectra are generated by projecting only the events along the
respective diagonal lines in Figure 2 onto the KER axis. Note that
for each value of n the ICD electron energy is given by
E$_{eICD}$(n)= (54.42 - 13.6 $\frac{4}{n^2})$ - 2 $\cdot$ 24.59 -
KER. The KER distribution displays a distinct oscillatory structure
which looks different for each n-state. In the classical reflection
approximation \cite{Gislason1973JCP} the KER of the coulomb
explosion is given by KER = 1/R, where R is the internuclear
distance at the instant of ICD. We have used this to obtain an
estimate of the internuclear distance at which ICD occurs (top panel
in Figure 1). Clearly this region of internuclear distances is much
more confined than the diffuse He$_2$ ground state. Only about 4 \%
of the ground state of He$_2$ is in the region 2 a.u. $<$ R $<$ 12
a.u. which we observe for ICD. As we will show below, more than 95
\% of the excited dimers decay via ICD. We suggest that significant
nuclear motion occurs prior to ICD. The He$^{*+}$(n=2,3...) - He
contracts until it reaches the regime where the ICD rates are
appreciable.

The ICD rate scales asymptotically like 1/R$^6$ \cite{Ced2002rep}.
Since this is a monotonic function the observed oscillatory
structure can only result from a structure in the R dependence of
the nuclear wavefunction before the decay
\cite{Santra00prl,moiseyev01jcp}. For n=2 we find 5 minima,
suggesting that vibrational wavefunctions with at least $\nu$=5 are
involved. Since the He$_2$ potential is much more shallow than the
He$^{*+}$(n=2,3...) - He potential energy curve, the neutral dimer
wave function is much more delocalized and its mean internuclear
distance is much larger. Hence the vertical transition by the
photoionization and excitation of He$_2$ populates mainly high lying
vibrational states of the He$^{*+}$(n=2,3...) - He. Since the
He$^{*+}$(n=2,3...) - He potential energy curves are attractive at
large distances, the vibrational wave packet will start to contract
\cite{Santra00prl,moiseyev01jcp}.

The idea that vibrational structure can become visible in ICD was
first suggested in pioneering theoretical work by Santra and
coworkers and Moiseyev et al. \cite{Santra00prl,moiseyev01jcp}. They
showed that whether or not a visible oscillatory structure in the
KER finally arises depends on the interplay of wave packet dynamics
and the R dependent decay rates. For Ne$_2$ such oscillation appear
in calculations only if unrealistic potential energy surfaces are
used \cite{ScheitJCP}. The He$_2$ system studied here is the only
system investigated so far which exhibits such structure under real
conditions. This is due to the unique delocalization of the neutral
ground state which results in the preferred population of the
excited states at very large distances. The high contrast of the
observed oscillations shows that either preferentially one high
lying vibrational state or only a few states where nodes in the wave
function coincide are populated.

For all previously studied species the ICD rates were such that,
whenever ICD was energetically allowed, it was faster than radiative
decay by several orders of magnitude \cite{Ced2002rep}. Due to the
1/R$^6$ scaling of the ICD rates, however, this is not true for the
large distances in the helium dimer. The lifetime for radiative
decay of He$^{*+}$(n=2 l=1) is 99.92 ps \cite{Drake1983}. Assuming
85 fs as a typical ICD lifetime at R = 6 a.u. \cite{Ced2002rep}, ICD
and radiative decay rates would become comparable at around R = 20
a.u.. In addition the attractive potential of the
He$^{*+}$(n=2,3...)-He at the large distances where it is populated
by the vertical transition is very shallow and hence classically the
dimer would contract very slowly. It is therefore not clear at all
what the overall branching ratio between ICD and radiative decay
will be in this system. To get an experimental estimate we have
searched for stable He$_2^+$ ions in our time-of-flight spectrum. In
coincidence with photoelectrons from ground state He$^+$(n=1) we
have found  a ratio of He$_2^+$ to He$^+$ monomers of about 2 \% as
expected from the dimer fraction in our beam. In coincidence with
electrons for He$^{*+}$(n=2), however, we did not find any stable
He$_2^+$ in our time-of-flight spectrum above the background. From
these numbers we conclude that the ratio of the excited
He$^{*+}$(n=2)-He decaying radiatively to the bound ground state of
He$_2^+$ to the ones which decay via ICD is  $<$5 $\%$. We note,
that this estimate was gained from an experiment at a photon energy
of 65,41eV which is only 1 meV above the n=2 threshold. This is
important since under these conditions the recoil energy of the
photoelectron imparted onto the He$^{*+}$(n=2)-He is only 68 neV and
we can hence safely exclude any significant influence of the recoil
effect on the ICD process \cite{Kreidi}.

In conclusion we have shown that a single photon leads to two-center
double ionization of the helium dimer in a two-step process. Firstly
one site is ionized and excited. This step is followed by ICD. The
kinetic energy release shows vibrational oscillatory structure from
the intermediate singly charged intermediate dimer state. Due to the
extreme condition ab initio calculations of decay rates and nuclear
dynamics are highly challenging. They are currently being performed
by the Cederbaum group and will be published separately.

\acknowledgments We want to thank the staff of BESSY for
experimental support. This work was supported by the Koselleck
Programm of DFG, R. E. G. acknowledges support by the Helmholtz
Society, grant VH-NG-331. We are grateful to L. S. Cederbaum, N.
Sisourat, N. Kryzhevoi and Ph. Demekhin for stimulating discussion.

\end{document}